\documentclass[fleqn,usenatbib]{mnras}
\usepackage{newtxtext,newtxmath}

\usepackage[T1]{fontenc}

\DeclareRobustCommand{\VAN}[3]{#2}
\let\VANthebibliography\thebibliography
\def\thebibliography{\DeclareRobustCommand{\VAN}[3]{##3}\VANthebibliography}

\usepackage{amsmath}
\usepackage{graphicx}

\pdfoutput=1

\makeatother
\usepackage{color}

\title[Nearly-zero CMB correlation] {Dipole-independent measurements of nearly-zero  CMB correlation: a possible symmetry of primordial causal quantum coherence}

\author[C. Hogan et al.]{
Craig Hogan$^{1}$\thanks{email: craighogan@uchicago.edu},
Ohkyung Kwon,$^{1}$
Stephan S. Meyer,$^{1}$
Nathaniel Selub,$^{1}$
and Frederick Wehlen$^{1}$
\\
$^{1}$University of Chicago, 5640 S. Ellis Ave., Chicago, IL 60637
} 

\date{Accepted XXX. Received YYY; in original form ZZZ}

\pubyear{\the\year{}}

\begin{document}
\label{firstpage}
\pagerange{\pageref{firstpage}--\pageref{lastpage}}
\maketitle

\begin{abstract}
Anisotropy of space-time is measured on the scale of the cosmic horizon, using the  angular correlation function $C(\Theta)$ of    cosmic microwave background (CMB) temperature at large angular separation $\Theta$.
Even-parity correlation  $C_{\rm even}(\Theta)$ is introduced to obtain a direct, precise measure of horizon-scale curvature anisotropy independent of the unknown dipole, with uncertainty  dominated by 
models of  Galactic emission.  In maps from {\sl WMAP} and
{\sl Planck}, $C_{\rm even}(\Theta)$ at $\Theta\simeq 90^\circ\pm 15^\circ$ is found to be much closer to zero than in previously documented measurements.  Variation from zero as small as that in the {\sl Planck} maps is estimated to occur by chance in a fraction  $\simeq 10^{-4.3}$ to $\simeq 10^{-2.8}$ of standard realizations.
 Measurements are found to be consistent with zero correlation in a range of angles expected  from  quantum fluctuations during inflation whose  spacelike coherence is bounded by inflationary horizons around every location at every epoch.  This scale-invariant  symmetry of cosmological initial conditions is
incompatible with the standard quantum theory of  initial conditions, but is broadly consistent with other cosmological  measurements, and is subject to further tests.
\end{abstract}

\section{Introduction}

The angular distribution of the cosmic microwave background  (CMB)  provides 
our most precise measurement of the  large-scale structure of space-time. 
The last scattering surface of the CMB lies at a comoving distance close to our causal horizon, and  the pattern of temperature on the sky on large angular scales is thought to represent a direct, intact relic of the large-scale distribution of space-time curvature in the initial conditions. On angular scales larger than a few degrees, the dynamics of the system and the propagation of light   are determined  only by  gravity \citep{Sachs1967,Bardeen1980,Hu:2001bc}. 
The CMB on such large angular scales is well known to display a precise symmetry: on average, it  is almost exactly  isotropic, characterized by the fact that at  large angular separations $\Theta$, its angular correlation function 
$C(\Theta)$, defined in Eqs. (\ref{harmonicsum}) and (\ref{skyaveragepoints}) below, has a dimensionless value much smaller than unity.

Both the near-perfect isotropy and the smaller-scale perturbations that lead to large scale cosmic structure can  in principle be generated by a causal physical process if there is an
early inflationary acceleration of the cosmological scale factor, which allows any two locations to have  causal contact at a sufficiently early time.  
In standard inflation theory,  perturbations arise from quantum fluctuations. The standard  model \citep{Baumann2009,Weinberg2008},  based on Gaussian noise derived from local effective quantum field theory (QFT), agrees with measurements of cosmic structure  over a wide range of scales, including CMB correlations on scales less than a few degrees.

However, this theory does not agree very well with the  precise  isotropy   measured on  larger angular scales.
Indeed, it has been realized since the earliest measurements of CMB anisotropy, with  the {\sl COBE} satellite \citep{Bennett1994,Hinshaw_1996}, that the universe is actually much smoother at  large $\Theta$ than the Gaussian model typically predicts.
The unexpectedly small magnitude of  $C(\Theta)$ at large angular separation was  confirmed with higher precision in subsequent  studies with  {\sl WMAP} and {\sl Planck} \citep{Spergel2003,WMAPanomalies,Planck2016,Planck-2018-Isotropy,Akrami:2019bkn}.

Even though  large-angle anisotropy provides the most direct probe of initial conditions,  correlations  at   $\Theta$ larger than a few degrees, or equivalently the angular power spectrum $C_\ell$ at angular wavenumber $\ell$ less than about thirty,  are often disregarded in tests of cosmological models, because the standard Gaussian model predicts many possible realizations of the CMB sky that differ significantly from each other on large angular scales. 
In standard cosmology, the small large-angle correlation is attributed to a statistical fluke of our particular realized sky; 
that is, only a small fraction of realizations that  agree with structure on smaller scales  are as smooth as the real sky on the largest scales. This persistent anomaly has led some authors to question the standard picture \citep{Copi2006,Copi2008,Copi2015,Muir2018}.

In spite of a theoretical expectation that casts doubt on its significance,
the structure of  CMB anisotropy on the largest  scales  nevertheless remains a  unique phenomenon,  which preserves a precisely measurable intact pattern of cosmic initial conditions.
Anomalously small large-angle anisotropy may be a direct signature of  physics that shaped  initial conditions.

Previous theoretical studies  have explored
 possible  fundamental implications of anomalous large-angle isotropy. In one notable study, \cite{Copi2018} studied toy models of primordial perturbations with compact support limited by a characteristic length,  based on  compact wavelets on 3D comoving spacelike hypersurfaces in initial conditions--- a real-space modification of the Gaussian noise   usually modeled by an initial QFT vacuum state defined in comoving harmonic 3-space.
They compare models with CMB maps and suggest  that the initial perturbations themselves may have a
suppressed correlation function on large length scales, which ``could signal that when inflationary perturbations are generated, they are coherent only over distances shorter than the horizon due to inflationary microphysics.''  

The measurements reported in this paper are designed to test a related  hypothesis about  microphysical constraints on inflationary correlations, 
formulated here as a covariant bound
in four dimensions
instead of a compact correlation in 3D. It is
based on a  physical principle  of causal coherence,  that quantum
systems in causally disconnected  regions of  space-time lead to independent outcomes.
In our application of this principle,   physical correlations of  perturbations from  every epoch during inflation are 
bounded by inflationary horizons  around every comoving location.
Such a causal bound is a natural consequence of the kind of
compact nonlocal causal spacelike coherence 
familiar in  physical systems with  ``spooky'' quantum correlations, but it  is incompatible with standard inflationary QFT formulations of gravitational fluctuations that depend on separability of space and time for their modal decomposition.\footnote{Causal coherence in 4D is also incompatible with  models that impose constraints on initial correlations  to be compact in comoving  real 3-space, as in the toy wavelet model of \cite{Copi2018}. As shown below, scale-invariant causally coherent correlations in 4D lead to nonzero 3D correlations, as well as exotic constraints on angular correlations,  at {\it all} comoving spacelike separations.} 
 We   show below how  covariant causal coherence in four dimensions can lead to angular correlations from primordial gravitational perturbations that are not only small, but  exactly vanish over a specific range of angular separation around $\Theta=90^\circ$.

The new measurements reported in  this paper are designed  to  test  this  angular symmetry. 
In  particular,  we introduce separate measurements of  even and odd parity correlations  to allow 
direct model-independent tests of exact universal null symmetries of angular correlation that do not occur in QFT models.


  The plan of this paper is as follows. 
  In Sec. \ref{motivation} we further explain the motivation and design of  the current study.  In  Sec. \ref{measurement} we  describe a new measurement of large-angle even-parity  CMB correlations.  In Sec. \ref{geometry}, we  review the relativistic causal structure of inflationary cosmology, and analyze  covariant   causal bounds in four dimensions that can lead to  relict symmetries of angular correlation.   In  Sec. \ref{comparison} we compare measurements with  standard QFT realizations at angular separations where  causal symmetries could lead to zero correlation.  In Sec. \ref{interpretation}, we summarize the alternative interpretations of the data.  An overall summary is presented in  Sec. \ref{conclusion}.
In the Appendix (Sec. \ref{appendix}), we describe in more detail how causally-coherent inflation   differs from the standard QFT picture, and give some examples of other ways it could modify standard concordance cosmology.

\begin{figure*}
\begin{centering}
\includegraphics[width=.9\linewidth]{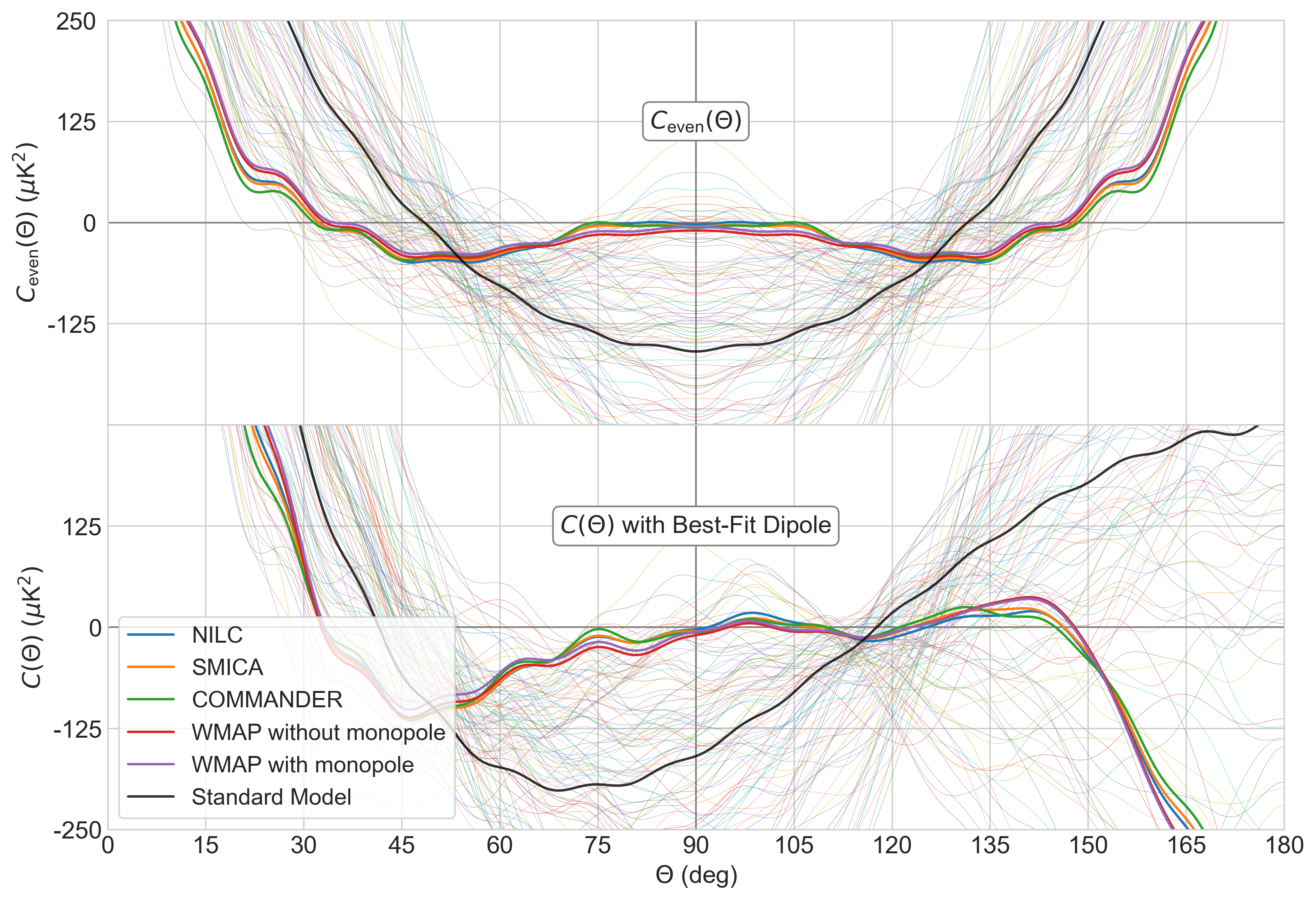}
\par\end{centering}
\protect\caption{Correlation functions of sky maps and standard-model realizations.  The top panel shows the  even parity angular correlation function of CMB temperature. Bold colors show $C_{\rm even}(\Theta)$ of Galaxy-subtracted all-sky maps from the {\sl WMAP}  and {\sl Planck} satellites, as labeled. For comparison, solid black shows the expectation of the standard model, and fine lines show 100 standard, randomly generated sky realizations. For causally coherent initial conditions,  this function should  vanish over a symmetric band  given by Eq.\;(\ref{evennull}), or $75.52^\circ<\Theta<104.48^\circ$,  which is strikingly approximated by the maps, especially those from {\sl Planck} data; for these, the variance of correlation is  three to four orders of magnitude smaller than that in typical standard realizations, as shown quantitatively in Fig.\;(\ref{cumulative}).
Bottom panel shows the total correlation with the best-fit dipole restored, assuming a causal shadow that extends over the maximal range tested, $75.52^\circ<\Theta<135^\circ$ (Eq.\;\ref{widetotalshadow}). 
For a fair comparison, in this panel each realization has a ``mock dipole'' correction added to minimize its departure from zero in the shadow region.  The maps approximate zero  more closely than almost any realization, even with this correction.}
\label{correlationspider}
\end{figure*}

\section{Motivation and design of this study}
\label{motivation}

The main goal of the current study is to use  CMB correlations on the largest scales to study possible exact symmetries of cosmological initial conditions.  This motivation drives  specific choices for measurements.

Systematic uncertainty in measurements of  large-angle CMB temperature correlations is mainly determined by 
the effect  of Galactic foregrounds.
A common practice for precision studies is to mask off large parts of the sky where models of the Galaxy are unreliable. This works well for some statistical tests, and with a fully defined Gaussian model such as standard inflation.  However, in general the unmeasured masked sections of sky generate biases that make them unsuitable for null tests \citep{Hagimoto_2020}. Instead, we adopt  
 all-sky models of the Galaxy 
developed by the {\sl Planck} and {\sl WMAP} teams using a variety of  different approaches, and use the differences between them as a guide to systematic uncertainties.


Another fundamental systematic measurement uncertainty arises  because CMB maps have unknown dipole ($\ell=1$) harmonic modes subtracted,  so that their measured $C(\Theta)$  differs from that of the CMB horizon itself \citep{Copi2018,Peebles2022}.
This uncertainty may be addressed in a parameter-free way by separate measurement of the even parity component of correlation $C_{\rm even}(\Theta)$, which has no dipole.
This measurement is  well suited to  test  a unique  symmetry-based prediction for  $C_{\rm even}(\Theta)=0$, where there is no particular theoretical expectation for the dipole.

The current work was partly motivated by   an earlier study   \citep{Hagimoto_2020}, which found that  the total $C(\Theta)$   at just one angle,
exactly $\Theta=90^\circ$, where contributions from the unknown dipole  identically vanish, lies in a range remarkably close to zero:
\begin{equation}\label{90correlation}
-0.22 \mu {\rm K}^2 < C(\Theta=90^\circ) < +2.16 \mu {\rm K}^2.
\end{equation}
That measurement  showed that the  CMB  at exactly 90 degrees is  hundreds of times smaller  than the value in typical standard realizations, 
and closer to zero than all but  $0.52\%$ of them. 

In this paper,  we extend this study to the whole even-parity part of the angular correlation function $C_{\rm even}(\Theta)$, which gives a direct estimate of true horizon-scale correlations,  independent of the unknown dipole or any other model parameter, over a wider range of angles. 
Our new measurements of  $C_{\rm even}(\Theta)$ are plotted for several maps in Fig. (\ref{correlationspider}), together with 100 random realizations of the standard model. 
 From the plot it can be immediately seen    that the absolute value of  $C_{\rm even}(\Theta)$ over a range of angles near 
$\Theta\simeq 90^\circ$ is still much smaller  than  expected, and   differs from zero  no more than  the different maps differ from each other.
Quantitatively, the surprising new  result derived below is that the variance of $C_{\rm even}(\Theta)$  from zero in the angular  range  $\Theta\simeq 90^\circ\pm 15^\circ$  is { \it  three to four orders of magnitude smaller than expected in the  standard cosmological model. }

The standard interpretation of this new fact, as before, is  that our particular horizon just represents a very unlikely statistical fluke, and its small correlations on such large scales are of no physical significance.
Such small correlation over a range of $\Theta$ is however  hard to dismiss lightly as a statistical anomaly: our rank comparison  shows that deviations from zero as small as those in the {\sl Planck} maps occurs in  standard realizations with probabilities that range from   $\simeq 10^{-4.3}$ to $\simeq 10^{-2.8}$, depending on the Galactic foreground model. 

A  small dimensionless number in nature can  often be traced to  a  fundamental symmetry.
We are bound to ask whether  the  measured near-perfect isotropy apparently preserved in  CMB correlation could possibly signify a fundamental symmetry of   initial perturbations that is not preserved in the  standard theory.
\footnote{Large-scale uniformity   ultimately depends on a symmetry of the initial state.
The actual measured large-angle uniformity is extraordinary  when expressed in absolute terms. To illustrate with one example from  estimates below,  the fractional  dimensionless correlation residual of the smoothest {\sl Planck}  map (NILC) around $\Theta\sim 90^\circ$, expressed as a fractional perturbation of total curvature  on the scale of the horizon, is $\int C_{\rm even}^2/(2.7{\rm K})^4\sim 2\times 10^{-26}$. In  standard theory, fluctuations are expected to add variance several orders of magnitude larger than this.}

A  fundamental symmetry that accounts for vanishing angular correlation would have to be a property of any sky, and any realization of initial conditions. It also needs to account for the specific range of angular scales, 
$\Theta\simeq 90^\circ\pm 15^\circ$, where nearly-vanishing even-parity correlation is observed. We show below that it is geometrically possible to formulate such a  symmetry,  defined by  covariant relativistic  geometrical relationships, that could account for large-angle $C(\Theta)$ measurements, and  at the same time  agree with the nearly scale-invariant  3D power spectrum of cosmological perturbations  on all  scales.

Our formulation is  based on  a  principle of  causal coherence widely tested in  entangled laboratory quantum systems \citep{Zeilinger1999,Vilasini2024}: namely, that   no correlation can occur between   systems contained within completely separate regions of space-time, because they are causally independent.
A  quantum process that occurs entirely in the future of one event, and entirely in the past of some future  event at the same location, can modify physical relationships only within  a unique 4D region of space-time bounded by their light cones, called a causal diamond. Quantum processes contained within completely separate causal diamonds do not produce physical correlations with each other.
A scale- and conformally-invariant symmetry of angular correlation in cosmic initial conditions could arise from physical potential differences generated on inflationary horizons from quantum fluctuations that are  completely contained within  causal diamonds during cosmic  inflation.

We show here that in principle, such a causally-coherent process  
 could  lead to an angular symmetry on any sky, at any time, similar to that measured in the CMB.
It is shown below that geometrically-derived  angular boundaries of causal correlations between world lines  during inflation coincide with  the range of angular separations  $\Theta\simeq 90^\circ\pm 15^\circ$ where we measure the smallest even-parity correlations.   The conformal causal relationships that lead to this result do not depend on  scale, consistent with a nearly scale-invariant spectrum in 3D.
In this range of angular separation, it is possible that causally-coherent primordial angular correlations are  not only  small in magnitude, they  actually vanish,  because  they are generated independently.
The simplicity of this geometrical symmetry makes it  possible to test in a model-independent way.

The angular  symmetry is formulated here from the standard conformal  geometry of  classical relativistic cosmology,  but it is not compatible with  the standard QFT model for quantum fluctuations and initial gravitational perturbations.  
The standard  formulation of initial conditions  has previously been challenged on theoretical grounds \citep{Penrose1989,Ellis1999,Ijjas2015},
and it is well known that  modifications of QFT are expected in a deeper theory of quantum gravity that addresses entanglement with causal horizons \citep{CohenKaplanNelson1999,HollandsWald2004,Stamp_2015}.
  If large-angle cosmic correlations  preserve unique information about nonlocal, causal superpositions of gravitational quantum states on horizons that do not occur in QFT, the exceptional 
symmetry of  CMB isotropy may be  more physically profound than it appears to be in the familiar context of  QFT-based inflation theory \citep{Hogan2019}.


\section{Measured large-angle anisotropy }
\label{measurement}

\subsection{Angular spectrum and correlation function}

In standard notation, the angular pattern of a quantity $Q$ on a sphere, such as scalar potential $\Phi$ or CMB temperature $T$,  can be decomposed into spherical harmonics
$Y_{\ell m}(\theta,\varphi)$:
\begin{equation}\label{decompose}
Q(\theta,\varphi)=
\sum_\ell \sum_m Y_{\ell m}(\theta,\varphi) a_{\ell m}.
\end{equation}
The harmonic coefficients $a_{\ell m}$ then determine
the angular power spectrum:
\begin{equation}\label{powerpiece}
C_\ell= \frac{1}{2\ell+1}
\sum_{m= -\ell}^{m=+\ell} | a_{\ell m}|^2.
\end{equation} 
The angular correlation function is given by its Legendre transform,
\begin{equation}\label{harmonicsum}
 C(\Theta) = \frac{1}{4\pi}\sum_\ell (2\ell +1) { C}_\ell P_\ell (\cos \Theta),
\end{equation}
where $P_\ell$ are Legendre polynomials.

As discussed below, $C(\Theta)$ can be separated into  two independent sums with  odd and even parity.  A sum with only even values of $\ell$ gives the unique even-parity correlation $C_{\rm even}(\Theta)$, which is symmetric around $\Theta=\pi/2$.

The same function can be expressed as an all-sky average
\begin{equation}\label{skyaveragepoints}
 C(\Theta)=\langle Q_1 Q_2\rangle_{\Theta}
\end{equation}
for all pairs of points $1,2$  separated by angle $\Theta$, or equivalently, an  average over all directions $\vec\Omega_i$
\begin{equation}\label{skyaverage}
 C(\Theta)=\langle Q_{\vec\Omega_i} \bar Q_{i\Theta}\rangle_{\vec\Omega_i}
\end{equation}
where $\bar Q_{i\Theta}$ denotes the  average value on a circle of angular radius $\Theta$  with center $\vec\Omega_i$. 

The power spectrum $C_\ell$ is the statistical tool generally used for tests of cosmological models.
However,  $C(\Theta)$  provides a more direct  signature  of geometrical causal relics of initial conditions, for reasons discussed below. 
Causal boundaries are not apparent in  $C_\ell$, even though it contains the same  statistical information about the angular distribution.


\subsection{Scalar perturbations and large-scale CMB anisotropy}

On large angular scales, temperature anisotropy in the CMB is mostly determined by primordial scalar curvature perturbations $\Phi$ of the cosmological metric on a thin sphere at the location of the last scattering surface \citep{Sachs1967}. Apart from the Doppler-induced dipolar anisotropy from local motion, their angular distributions on scales larger than a few degrees are the same:
\begin{equation}
\delta T(\theta,\varphi)\propto \Phi(\theta,\varphi), 
\end{equation}
and therefore so are their angular correlations:
\begin{equation}\label{TandPhi}
    C_T(\Theta)\propto C_\Phi(\Theta).
\end{equation}
In this sense, CMB correlation provides a direct measurement of any angular symmetry of initial conditions on a particular sphere.

Gravity also introduces some anisotropy during propagation, via the integrated Sachs-Wolfe effect (ISW) \citep{Hu:2001bc,Francis2010,Copi2016}. This effect  is generated by primordial perturbations in 3D, as the CMB light propagates through space on our past light cone. In the linear regime, it is also determined by the  invariant local scalar potential $\Phi$ that preserves its  original primordial spatial distribution from the end of inflation. Anisotropy from this effect comes from perturbations at comoving distances  smaller than the last scattering surface. Thus, symmetries of CMB  temperature correlation are mainly determined by the SW effect of $\Phi$ on the last scattering surface (Eq. \ref{TandPhi}), with a relatively small additional ISW contribution shaped by  
angular cross-correlations of $\Phi$ in three comoving spatial dimensions \citep{Copi2016}.


In the analysis below, we will neglect other physical effects, such as radiation transport and Doppler motion at recombination, which do not modify the angular spectrum significantly at spherical harmonics with $\ell\lesssim 30$ \citep{Hu:2001bc}.

\subsection{Data}
\label{sec:data}
As explained in \cite{Hagimoto_2020}, we use all-sky CMB maps made with subtracted models of Galactic emission, in order to minimize correlation artifacts introduced by masks. Our analysis is based on foreground-corrected maps of the CMB temperature based on the fifth and third public release databases of the \textsl{WMAP} and \textsl{Planck} collaborations, respectively. For \textsl{WMAP}, we use the \textsl{ILC} map with and without the fitted monopole included\footnote{Both of these possibilities were presented as models by the \textsl{WMAP} team, so we show them here for completeness. In fact, this monopole must be interpreted as an artifact of imperfect Galaxy model subtraction: there can be no actual measured monopole of true CMB anisotropy by definition, since the monopole is isotropic.}. In the case of \textsl{Planck}, we use several different maps based on different techniques for modeling the Galaxy. Recognizing that the noise properties of the foreground-corrected maps are not well characterized and that the 2-point function is correlated between angles, we use the variation between foreground subtraction methods and experiments as a proxy for correlation function uncertainty. We only compare integrated residuals of measured values and standard model realizations of 2-point correlation functions.

For this paper, we used the python wrapper for the Hierarchical Equal Area isoLatitude Pixelization (\textsl{HEALPix}) scheme \citep{Gorski_2005} on maps at a resolution defined by $N_{\text{side}} = 256$. We preprocessed the maps by converting them to this resolution and removing their respective dipole spherical harmonic moments.  We conduct all measurements and operations on each map independently.

\subsection{Measured correlation function}
The top panel of Fig.\;(\ref{correlationspider}) shows our main new result: a direct measurement of  even-parity CMB angular correlation, which is independent of the unknown dipole. 
It reveals a simple  fact, that  the absolute value of $C_{\rm even}(\Theta)$ over a significant range of angles is remarkably close to zero.

 In particular, the  measured variation of $C_{\rm even}(\Theta)$ from zero in the range $\Theta\simeq 90^\circ\pm 15^\circ$ is orders of magnitude smaller than previously documented correlations.
Its absolute value   is comparable with differences between  different foreground-subtracted maps, that is, apparently as close to zero as current measurements allow.
All three {\sl Planck } maps--- which  arguably provide the most accurate models of Galactic foregrounds--- are nearly indistinguishable from  zero on the scale plotted in Fig.\;(\ref{correlationspider}).

The anisotropy  is also strikingly small when compared to the expectations of standard cosmology.  To illustrate this comparison, Fig.\;(\ref{correlationspider}) shows 100 examples of $C_{\rm even}(\Theta)$ produced in standard realizations--- that is, anisotropy produced from  the same quantum fluctuations that generate  cosmic structure on smaller scales.

We now proceed to  describe how the observed nearly-zero  $C_{\rm even}(\Theta)$ could be interpreted as a  causal symmetry of causal initial conditions, and then to quantitative comparisons of this interpretation with the standard picture.

\begin{figure*}
\begin{centering} \includegraphics[width=.9\textwidth]{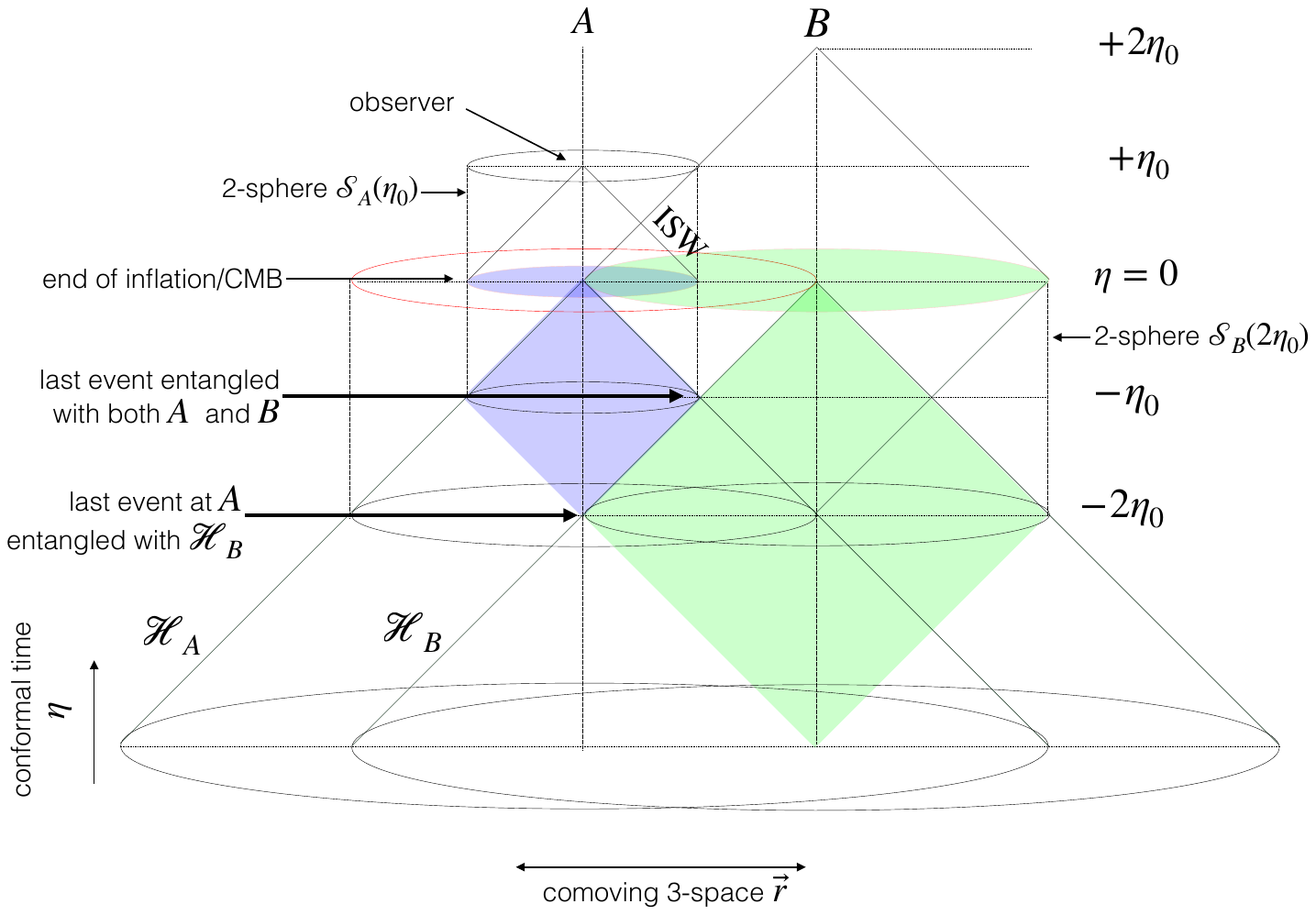}
\par\end{centering}
\protect\caption{
Four dimensional spacetime in  conformal coordinates, showing causal relationships  between  world lines $A$ and $B$ spaced a comoving distance $2\eta_0$ apart.
 While the relationships in the figure hold in general for any two world lines and any scale, to analyze the situation of the observation of the CMB, we imagine the observer to be on world line $A$ at a time $+\eta_0$ and set the time $\eta = 0$ to be the end of inflation and shortly after, at the recombination epoch. Thus $\eta_0$ is the current horizon distance and the CMB is emitted by a 3-sphere with radius $\eta_0$,  denoted $\mathcal{S}_A(\eta_0)$.
 The goal is to analyze what events during inflation can, even in principle, generate correlations in the CMB and determine the pattern of the correlations.
 CMB temperature anisotropy on large angular scales is dominated by gravitational perturbations near the last scattering surface, near ${\cal S}_A(\eta_0)$, but also has contributions from distortions arising from later perturbations along the light cone at $\eta > 0$, shown as ISW in the figure. Incoming information to the world lines $A$ and $B$ during inflation is bounded by inflationary horizons ${\cal H}_A$ and ${\cal H}_B$.
 Shaded 4D regions represent causal diamonds  bounded by the comoving 2-spheres ${\cal S}_A(\eta_0)$ (blue) and ${\cal S}_B(2\eta_0)$ (green). We posit that coherent quantum fluctuations are bounded by causal diamonds and convert into classical potential differences on inflationary horizons, so a coherent perturbation of  ${\cal S}_A(\eta_0)$ forms within the blue causal diamond shown, which starts on $A$ at  $-2\eta_0$; this is the last causal diamond  whose fluctuations  entangle $A$  with $B$.  
 At locations $\vec r$ with radius $|\vec r-\vec r_A|<\eta_0$,  correlations of  $\Phi(\vec r)$   with  $\Phi(\vec r_A)$  are  independent of correlations of  $\Phi(\vec r_A)$  with $\Phi(\vec r_B)$, so the intersection of  spheres ${\cal S}_A(\eta_0)$ and ${\cal S}_B(2\eta_0)$ represents the   boundary of  nonzero correlation measured at $(\vec r_A, \eta_0)$. A $B$ horizon at larger distance intersects ${\cal S}_A(\eta_0)$ at a larger angle, but its fluctuations are not entangled with ${\cal S}_A(\eta_0)$.  This leads to a maximum angular separation for causal correlation,  shown in Fig. (\ref{3Dshadow}).}
\label{nested}
\end{figure*}


\section{Causal correlations during inflation}
\label{geometry}

\subsection{Conformal causal structure}
The   metric for any homogeneous and isotropic cosmological space-time can be written in conformal coordinates 
 \citep{Baumann2009,Weinberg2008} as
\begin{equation}\label{FLRW}
    ds^2 = a^2(t) [c^2d\eta^2- d\Sigma^2],
\end{equation}
where $t$ denotes proper cosmic time for any comoving observer, $d\eta\equiv dt/ a(t)$ denotes a conformal time interval, and $a(t)$ denotes the cosmic scale factor.  For a  spatially flat model like that observed, the spatial 3-metric in comoving coordinates is
\begin{equation}\label{flatspace}
    d\Sigma^2 = dr^2 + r^2 d\Omega^2,
\end{equation}
where $r$ is the comoving radial coordinate, and the angular separation $d\Omega$ in standard polar notation satisfies $d\Omega^2 = d\theta^2 + \sin^2 \theta d\varphi^2$. Light cones and causal diamonds are defined by null relationships in comoving conformal coordinates,
\begin{equation}\label{null}
    d\Sigma = \pm cd\eta.
\end{equation}
Thus, in conformal coordinates, cosmological causal relationships throughout and after inflation are the same as those in flat spacetime.
 We adopt coordinates where $\eta=0$ corresponds to the end of inflationary acceleration. On the large scales studied here, it can also be identified with the CMB last scattering surface.
In the following, we set $c=1$.
Some key relationships and causal diamonds in this geometry are illustrated in Figure (\ref{nested}).

\subsection{Causal bounds on coherence and correlation}

Cosmological inflation \citep{Baumann2009} was introduced to solve a conceptual problem with initial conditions in classical cosmology, sometimes called the ``horizon problem'': as the cosmic expansion slows with time due to normal gravity ($\ddot a<0$), causal connections are only possible over smaller comoving regions in the past, so there is no causal mechanism for generating any kind of correlations in the initial conditions.

Inflation solves the main problem by introducing early cosmic acceleration, so that the comoving causal horizon moves closer with time rather than farther away
\footnote{An excellent tutorial including visualizations of inflationary conformal space-time, causal horizons, and  sky projections  can be found at \url{https://www.astro.ucla.edu/~wright/cosmo_04.htm}}.  If the scale factor $a(t)$ undergoes many orders of magnitude of expansion during early acceleration with $\ddot a>0$ before some epoch $\eta=0$, even very distant comoving world lines were once in causal contact. This causal relationship is shown in Fig.\;(\ref{nested}):  a comoving world line at any  finite radial distance  $\eta$ lies within the past light cone  or  ``inflationary horizon'' ${\cal H}$ at times earlier than $-\eta$.
Fig.\;(\ref{nested}) shows  spatial ``footprints'' of horizons: comoving spherical  surfaces ${\cal S}(\eta)$ that pass through the horizon and out of causal contact at time $-\eta$, and come back into view after inflation at time $+\eta$. All points on a spherical surface ${\cal S}(\eta)$ have a  causal connection with an event at its center at $-\eta$, so  large-scale homogeneity and isotropy, as assumed in Eq. (\ref{FLRW}), can in principle be generated by a causal physical process. 

Now consider the  physical process that generates spatial departures from uniformity caused by quantum fluctuations. In the analysis below, we consider a new causal symmetry that follows from a stronger causal constraint than  standard inflationary theory. Suppose that  all gravitational quantum fluctuations are causally coherent, in a sense well established from direct measurements of quantum entanglement \citep{Zeilinger1999,Vilasini2024}: 
physical  correlations  are not generated by systems in  separate regions of space-time.
This constraint requires that  physical effects of coherent quantum fluctuation states are spatially compact.
Specifically, we posit that {\it quantum curvature fluctuations on any world line interval are entangled nonlocally with other locations only
within  the compact 4-dimensional region of space-time encompassed by its causal diamond.}
Put another way, physical correlations  are  bounded by two-way causal relationships.
\footnote{Effects of causally coherent fluctuations with the same physical origin in flat space-time may be detectable in proposed laboratory experiments
\citep{vermeulen2020,Kwon2022,vermeulen2024photon}.
}

We further posit that the inflationary horizon  ${\cal H}$  imprints a sharp boundary on  coherent quantum fluctuations that create  curvature perturbations correlated with that world line: 
   {\it Quantum fluctuations  create differences in classical potential from any world line in the  future of its inflationary horizon ${\cal H}$}. That is, the classical potential difference between world lines forms when they cross each others' horizons.

As discussed in the Appendix, this  hypothesis about  how quantum fluctuations convert into classical perturbations  differs physically from the  freezing of fluctuations in the standard QFT model, where coherent  plane waves freeze  independently on each comoving scale by synchronous cooling  as their wavelengths stretch beyond the horizon scale. In that picture, the final observed classical correlations on ${\cal S}(\eta)$ are fixed by initial data laid down coherently in a region much larger than $\eta$,  at a time much  earlier than $-\eta$. 
Perturbations in the standard picture are defined in relation to a fixed initial background; in a causally coherent model, they are defined relationally between world lines, as allowed by causal relationships in a coherent quantum system. 
In standard inflation, the final outcome everywhere is fixed by state of the initial vacuum at the start of inflation; in a causally coherent picture, the quantum state and the final relational perturbations remain in a superposition within horizons until the end of inflation.


Our formulation is conformally invariant, so the coherent causal constraint applies to relational perturbations on all comoving length scales. As discussed in the Appendix, there may be observable effects of exotic high-order correlations in the 3D pattern of classical curvature perturbations on smaller scales than the current horizon.

\begin{figure}
\begin{centering} \includegraphics[width=\linewidth]{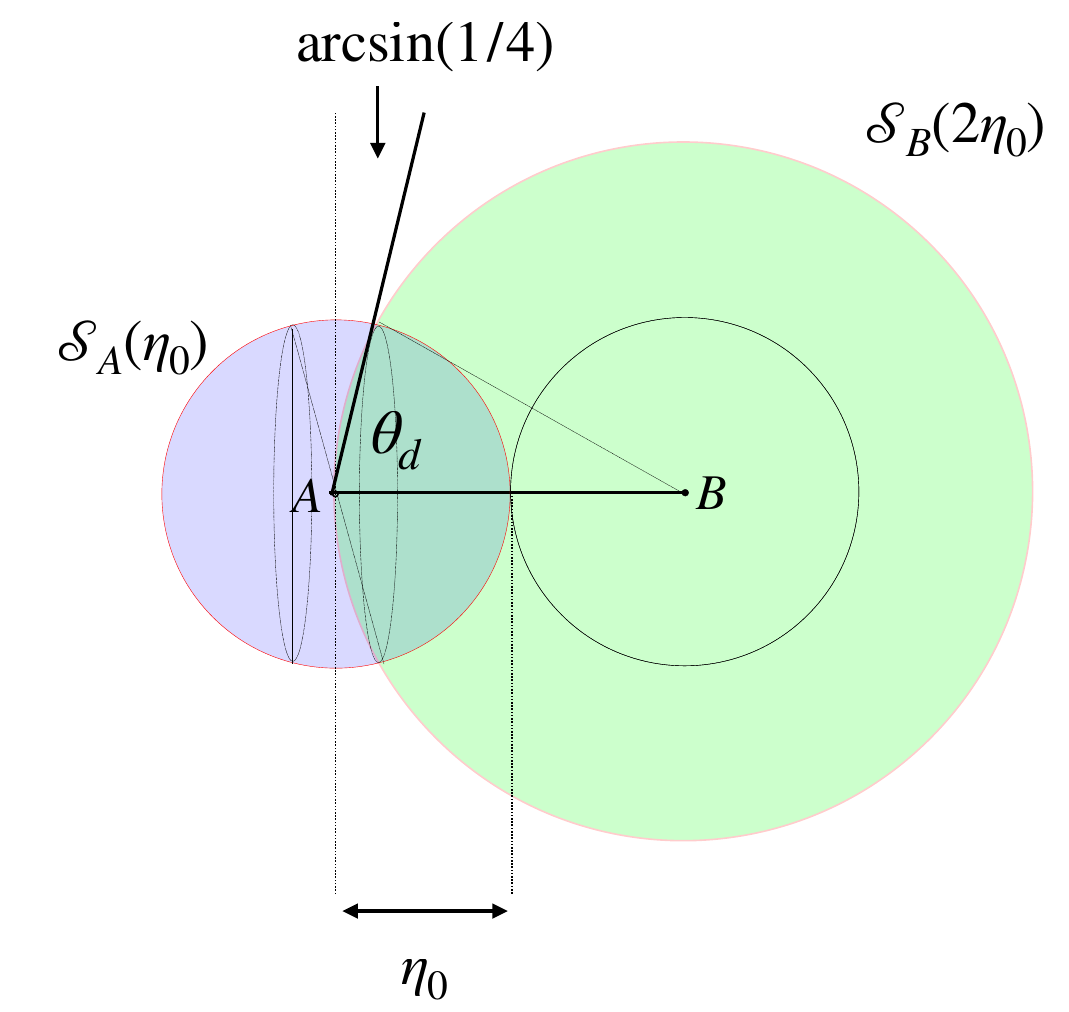}
\par\end{centering}
\protect\caption{A section of relationships in Figure \ref{nested}  at time $\eta=0$. The blue and green disks represent equatorial slices of comoving spheres at $\eta = 0$, shown as ellipses in Figure \ref{nested}. This projection illustrates the geometrical derivation of the disentanglement angle  (Eq. \ref{disentanglement}). The angular radius of the intersection circle on ${\cal S}_A(\eta_0)$,  $\theta_d= \pi/2-\arcsin(1/4)$, is the maximum  angular separation for correlation  with  any  direction, so  the directional average in the correlation function $C(\Theta)$ (Eq. \ref{skyaveragepotential})  vanishes at $\Theta=  90^\circ\pm \arcsin(1/4)$  (Eq. \ref{3Dnull}). At the same time, perturbation differences from $\Phi_B$ on ${\cal S}_B(2\eta_0)$ are independent of causal restrictions, as in the standard picture, at angular separations  $\lesssim 2\arcsin(1/4)\simeq 29^\circ$. These angular causal relationships are conformally invariant, so they apply to any location, epoch or comoving length scale $\eta_0$.}
\label{3Dshadow}
\end{figure}

\subsection{Scale-invariant angular boundaries of  causal correlation}


\subsubsection{Zero correlation around $\Theta= 90^\circ$ from disentanglement}


As discussed above, suppose that  causal correlations are bounded by causal diamonds and perturbations form on inflationary horizons. 
Fig.\;(\ref{nested})  shows causal diamonds around two world lines $A$ and $B$, with separation $2\eta_0$.
The causal diamonds on these world lines that begin after 
$-2\eta_{0}$ are disentangled from the other: their fluctuations constitute separate quantum systems, and   generate uncorrelated perturbations on their respective horizons.

These world lines are shown because the circular intersection  of ${\cal S}_A(\eta_0)$ with ${\cal S}_B(2\eta_0)$ 
is the largest  angle from the $AB$ axis where $\Phi$ at $|\vec r|\leq \eta_0$ is  correlated by entanglement with horizons centered at  any location on  the axis.
(Horizons of points $B$ at larger distances from 
$A$ intersect at larger angles,  but they are not entangled within ${\cal S}_A(\eta_0)$.)


Slices of the corresponding comoving causal-diamond boundary surfaces in 3D are shown in Fig.\;(\ref{3Dshadow}). The 
angular  radius   on ${\cal S}_A(\eta_0)$
of its intersection with  ${\cal S}_B(2\eta_0)$ is 
the disentanglement angle,
\begin{equation}\label{disentanglement}
 \theta_d =\pi/2-\arcsin(1/4)\simeq 75.52^\circ.
\end{equation}
As viewed from $A$ at time $\eta_0$,  perturbations    separated from the $B$ direction $\vec \Omega_B$
by more  than $\theta_d$ are independent of those at $B$.


The angular correlation function  is given by Eq.\;(\ref{skyaverage}), an all-sky  correlation with circles of radius $\Theta$ centered on every direction $\vec \Omega_B$:
 \begin{equation}\label{skyaveragepotential}
 C(\Theta)=\langle \Phi_{\vec \Omega_B} \bar \Phi_{\vec \Omega_B\Theta}\rangle_{\vec \Omega_B}.
\end{equation}
Independence of $\Phi$ at polar separation exceeding the disentanglement angle (Eq. \ref{disentanglement}) from every direction  $\vec \Omega_B$ then leads  to  zero  correlation above the disentanglement angle,
\begin{equation}\label{totalzero}
    C(\Theta>\theta_d)=0.
\end{equation}

Importantly for direct null tests of CMB anisotropy,  the bound includes angular  cross correlations at different radii (of  $\Phi_{\vec\Omega_B}(|\vec r|<\eta_0)$ with $\bar\Phi_{\vec \Omega_B\Theta} (|\vec r|=\eta_0)$), so it 
applies to gravitational anisotropy  generated  via the ISW effect as well as  from last scattering.


The even- and odd- parity components of $C(\Theta$  are respectively symmetric and antisymmetric around $\Theta=\pi/2$,  so 
 they must both have zero correlation over a range of angles
symmetric around $\pi/2$:
\begin{equation}\label{3Dnull}
C_{\mathrm{total}}=
C_{\mathrm{even}}=C_{\mathrm{odd}}=0
\ \qquad
( 75.52^\circ<\Theta<104.48^\circ ).
\end{equation}
As described below, because odd parity correlations always vanish at exactly $90^\circ$ and the dipole component is not measured, the even-parity component 
provides the most powerful direct test of this ``causal shadow'' symmetry.


 \subsubsection{Correlation at larger angular separation}

 At angles larger than $\pi/2+\arcsin(1/4)\simeq 104.48^\circ$, 
tests of null symmetry (Eq. \ref{totalzero}) 
 must include   odd-parity correlation.
Moreover, where  odd and even parity components do not both vanish, subtraction of unmeasured monopole and dipole components  can indirectly generate apparent angular correlations outside the symmetric interval of the 3D causal shadow (Eq. \ref{3Dnull}). \footnote{Apparent large-angle correlation introduced by  monopole and dipole subtraction was studied in the context of  wavelet models by \cite{Copi2018}.}


We have not derived  a model-independent  causal  symmetry for angular separations outside  the 3D causal shadow.
Even so,  it is interesting to explore empirical constraints on  correlations that include both  odd and even parity components, and that also account for the unobserved dipole, to test whether data is  consistent with zero total true correlation over a  larger range of angular separations. We choose to test two ranges with possible geometrical origins.
For one test,  we will exclude separations within  the intersection angle of a horizon of equal radius centered on the antipodal point, $\Theta>2\pi/3$:
\begin{equation}\label{generaltotalshadow}
 C_{\rm minimal}([\pi/2-\arcsin(1/4)]<\Theta<2\pi/3)
 =0.
\end{equation}
We will also test the possibility of a wider causal shadow that could be generated by causally-coherent ``tilted'' perturbations: 
\begin{equation}\label{widetotalshadow}
 C_{\rm maximal}([\pi/2-\arcsin(1/4)]<\Theta<3\pi/4)
=0.
\end{equation}

\section{Tests of CMB Symmetries }
\label{comparison}

\subsection{Dipole subtraction and parity separation}

It is not possible to measure the true primordial pattern in the CMB,  because the dipole components $a_{1m}$ have been removed from the maps to compensate for the local motion relative to the local cosmic rest frame, including our nonlinear orbits within the galaxy and the Local Group. These motions are not known to nearly enough precision to separate the primordial dipole \citep{Peebles2022}. 
 Nevertheless, a small fraction of the subtracted dipole is part of the intrinsic large-angle primordial pattern on spherical causal diamond surfaces and contributes to correlation in the angular range of causal shadows. Thus, a null shadow symmetry can only become apparent when the intrinsic portion of the dipole is included. If it is a true symmetry of  gravitational potential correlation, then there must exist a dipole that can be added to the observed CMB temperature map that realizes the  symmetry.
 \footnote{We omit consideration of second-order Doppler anisotropy, which generates even-parity harmonics including a quadrupole, but with amplitude smaller than our measurement precision.}

The total correlation (Eq.\;\ref{harmonicsum}) is a sum of even and odd Legendre polynomials, which are respectively symmetric and antisymmetric about
$\Theta=\pi/2$. To produce zero correlation over a range symmetric around $\Theta=\pi/2$, no combination of even functions can cancel any combination of odd ones, so if an angular correlation function vanishes over a range $[\pi / 2 - \Theta_0, \pi / 2 + \Theta_0]$ symmetric about $\Theta = \pi / 2$, the even contributions and the odd contributions to the angular correlation function must vanish independently over that range.

This property allows a direct, model- and dipole- independent test of causal symmetry, that uses only even-parity correlation.
In  a band of angles symmetric around $\Theta=\pi/2$  determined by the  causal shadow
(Eq.\;\ref{3Dnull}),
the sum of even terms must vanish on its own, independently of any dipole or model parameters.
The  causal shadow of even-parity correlation is thus
\begin{equation}\label{evennull}
C_{\rm even}(|\Theta-\pi/2|<\arcsin[1/4])=0,
\end{equation} 
or approximately
$C_{\rm even}(75.52^\circ<\Theta<104.48^\circ)=0$.

Furthermore, if the true primordial angular correlation including $\ell = 1$ vanishes over an arbitrary range $[\alpha, \beta]$, then the sum of the even and odd Legendre polynomials  in Eq. (\ref{harmonicsum}), measured only with  $\ell > 1$, departs from zero by a  function of known form, the dipole harmonic term
\begin{equation}\label{dipolecosine}
 \mathcal{D}(\Theta)=
 C_{\rm dipole\  only}(\Theta)= \frac{3}{4\pi}C_1 \cos(\Theta),
\end{equation}
where $C_1 \geq 0$. Thus, if there is a causal shadow  over a larger   angular range (Eq.\;\ref{generaltotalshadow}~or~\ref{widetotalshadow}), the sum of the even and odd Legendre polynomials for $\ell > 1$ must vanish after addition of a dipole of unknown amplitude.

\subsection{Comparison with standard predictions}

\subsubsection{Even-parity and total correlation comparisons}

We perform two types of comparisons with  data. First, we directly compare the maps with the zero correlation predicted from causal coherence in the  3D causal shadow, using only even-parity correlation.  Then, we  use  model-independent comparisons of data  to explore whether maps are also consistent with zero total correlation over a larger range of angles, where  the unmeasured dipole must be accounted for.

According to the causal shadow hypothesis,  the even-parity  correlation should vanish in the range of angles where all gravitational contributions  to both odd and even contributions vanish 
(Eq. \ref{evennull}).
As verified quantitatively by the rank comparison described below, the maps are indeed much closer to zero over this range than almost all realizations in the standard picture. 
Prediction and measurement in this comparison are model- and parameter-free.

Outside this range, the odd and even components do not separately vanish. The unmeasured dipole must be included to reveal any null symmetry, since odd-parity harmonics must  be included. 
An added cosine function (Eq.\;\ref{dipolecosine})  reproduces the effect of restoring any unobserved intrinsic dipole. The amplitude of this function is not known, which  must be accounted for in statistical comparisons.

\subsubsection{Standard realizations}

To generate standard-model realizations, we used the {\sl Code for Anisotropies in the Microwave Background} ({\sl CAMB}) \citep{CAMB2011} to calculate $C_\ell^\text{\,SM}$ with the following six cosmological parameters from the \textsl{Planck} collaboration \citep{2020A&A...641A...6P}: dark matter density $\Omega_c h^2 = 0.120$; baryon density $\Omega_b h^2 = 0.0224$; Hubble constant $H_0 = 67.3$; reionization optical depth $\tau = 0.054$; neutrino mass $m_\nu = 0.06$ eV; and spatial curvature $\Omega_k = 0.001$. For each  realization, we calculated the angular power spectrum  {using} Eq.\;\ref{powerpiece}. Then, we determined $C(\Theta)$ by summing Eq.\;\ref{harmonicsum} up to the sharp cutoff {at} $\ell_{\text{max}} = 30$.

Correlation functions of realizations and CMB maps are shown in Fig.\;(\ref{correlationspider}). On this scale,  realizations with the same parameters display  considerable cosmic variance. 
The 100 realizations shown for $C_{\rm even}$ directly illustrate examples of what would be expected in the standard picture.

Standard realized correlation functions include only $
\ell>1$ harmonics. For the comparisons of total correlation shown in the second panel of Fig.\;(\ref{correlationspider}), which include odd-parity harmonics, each realization is modified with a function of the form in Eq.\;(\ref{dipolecosine}) to minimize its residuals from zero. For realizations, this term does not have any relation to an actual physical dipole: it is a ``mock dipole'' added to estimate how frequently the sum of $\ell>1$ harmonics in Eq.\;(\ref{harmonicsum}) comes as close to the maps as zero correlation in the posited range of angular separation,  even if a dipole of unrestricted value is included.\footnote{Our likelihood estimates are conservatively generous to the standard picture; we have not allowed for the fact that most realizations do not have dipoles as large as the mock dipoles.}
Although the residual variance of these comparisons exceeds that of  purely even parity correlation in the 3D causal shadow, it appears that the measured $C(\Theta)$ in the posited range (Eqs.\;\ref{generaltotalshadow}~or~\ref{widetotalshadow}) is still closer to zero than almost all standard realizations, even when they have a mock dipole added. 

\subsubsection{Residuals}
The striking visual impression of a  null symmetry in the measured correlation can be verified quantitatively by a rank comparison of residuals.
For angular power spectrum $C_{\ell}$, define the even-parity angular correlation function $C_{\text{\rm even}}(\Theta)$ as
\begin{equation}
    C_{\text{\rm even}}(\Theta) = \frac{1}{4\pi} \sum_{\substack{\ell\;\!=\;\!2,\;\!4,\;\!6,\;\!...}}^{\ell_{\text{max}}}(2\ell +1) { C}_\ell P_\ell (\cos \Theta),
    \label{eq:ceven}
\end{equation}
where $\ell_{\text{max}} = 30$. Let $\{\Theta_{j, [\alpha, \beta]}\}_{i = 1}^N$ denote a uniformly spaced lattice of points in the range $[\alpha, \beta]$. Then, define the even-parity residual
\begin{align}
\Delta_{\text{\rm even}, [\alpha, \beta]}(C(\Theta)) &\equiv \int_\alpha^\beta|C_{\text{\rm even}}(\Theta)|^2 \, d\Theta \:\! \\
& \hspace{-30pt}\approx \sum_{i = 1}^N [C_{\text{\rm even}}(\Theta_{i, [\alpha, \beta]})]^2 \cdot \left(\frac{\beta - \alpha}{N}\right)\!
\label{eq:deltaEven}.
\end{align}
Similarly, for the total residual, we define
\begin{align}
\Delta_{\text{best-fit}, [\alpha, \beta]}(C(\Theta)) &\equiv \int_\alpha^\beta|\tilde C_{\beta}(\Theta)|^2 \, d\Theta \:\! \\
 & \hspace{-30pt}\approx \sum_{i = 1}^N [\tilde C_\beta(\Theta_{i, [\alpha, \beta]})]^2 \cdot \left(\frac{\beta - \alpha}{N}\right)\!,
\label{eq:deltaBestFit}
\end{align}
where
\[
\tilde C_\beta = C + \mathcal{D}_{\text{best-fit}}
\]
and $\mathcal{D}_{\text{best-fit}}$ is the dipole contribution that minimizes the residual.

We then define  integrated variation residuals over three ranges of angles:
\begin{align}
    \Delta_{\text{\rm even}} &\equiv \Delta_{\text{\rm even}, [\pi/2 - \arcsin(1/4),\ \pi/2 + \arcsin(1/4)]},\\
    \Delta_{\text{minimal}} &\equiv \Delta_{\text{best-fit}, [\pi/2 - \arcsin(1/4),\ 2\pi / 3]},\\   
    \Delta_{\text{maximal}} &\equiv \Delta_{\text{best-fit}, [\pi/2 - \arcsin(1/4),\ 3\pi / 4]}.
\end{align}
using the angular relationships discussed above (Fig. \ref{3Dshadow}).
We use these three residuals as a measure of how compatible a given power spectrum $\left\{C_{\ell}\right\}_{\ell > 1}$ is with the causal shadow symmetry. Each of the integrals must vanish for  a power spectrum that exactly agrees with the causal shadow symmetry in the specified range. In practice, we found that $N = 2000$ is a sufficiently high lattice resolution to approximate the integrals among different data sets and standard model realizations with negligible error. 

\begin{figure*}
\begin{centering}
 \includegraphics[width=\textwidth]{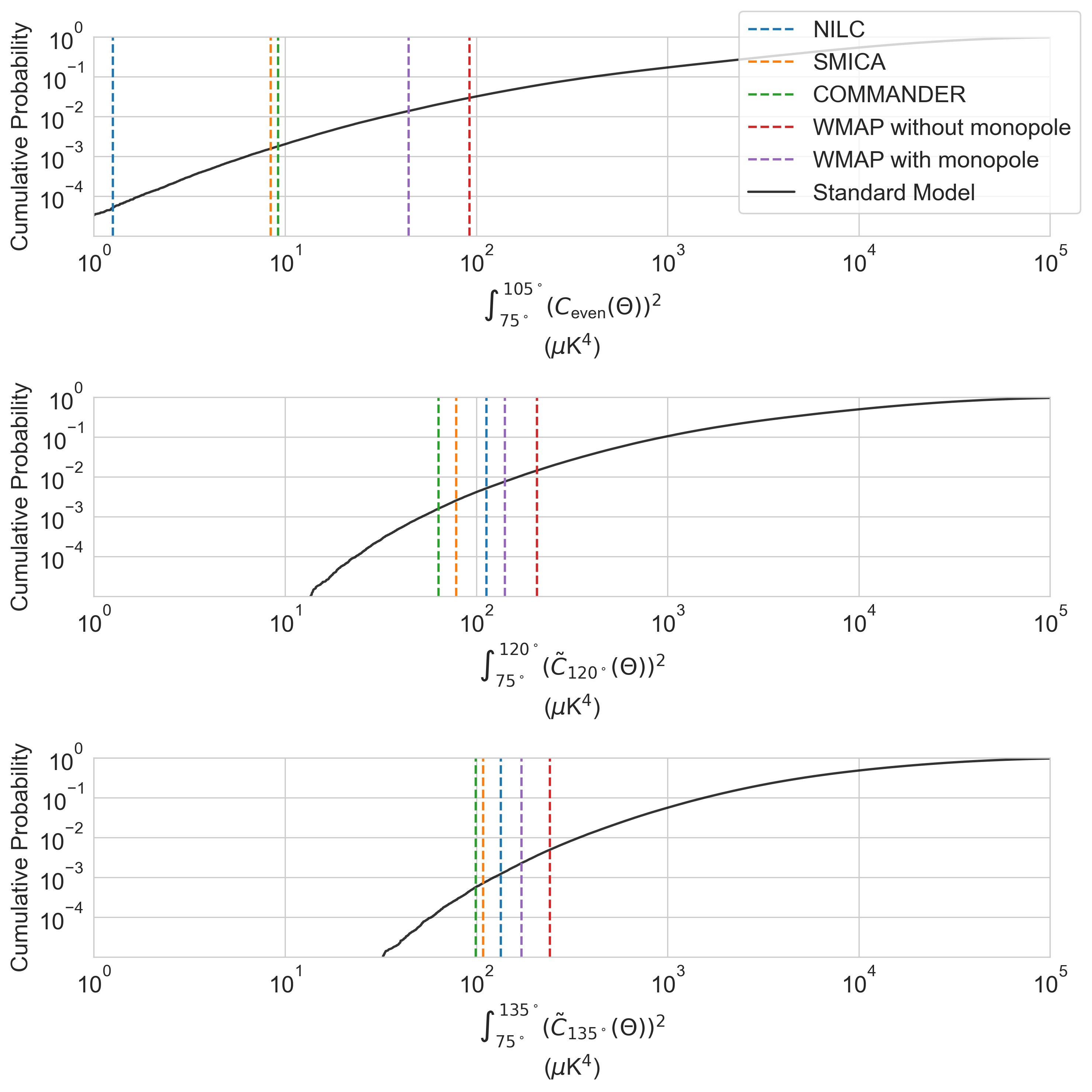}
\par\end{centering}
\protect\caption{Cumulative probability of deviations from zero correlation in standard realizations, compared with deviations of CMB maps, over various ranges of angular separation. The top panel directly compares  deviations of even-parity correlation (Eq.  \ref{eq:deltaEven}) over the computed 3D causal shadow (Eq.\;\ref{evennull}). No parameters are used for this comparison. The middle panel compares deviations  of total correlation (Eq. \ref{eq:deltaBestFit}) over the minimal asymmetric shadow (Eq.\;\ref{generaltotalshadow}), 
and the bottom panel compares deviations of total correlation over the maximal asymmetric shadow (Eq.\;\ref{widetotalshadow}); both of these allow for a mock-dipole correction to minimize residuals for each realization. 
In spite of variation between the maps, their variations  are all much closer to zero than almost all standard realizations. The  departures from zero are  comparable with the differences between the maps, as expected if they are dominated by systematic measurement errors. In all of these comparisons,  the {\sl Planck} maps match zero  better than the {\sl WMAP} maps. In the cleanest direct comparison, which is represented by the top panel,  the  residuals $\Delta_{\text{\rm even}}$ found in  the  {\sl Planck} maps are only 1 to 9 $\mu{\rm K}^4$, compared with $\Delta_{\text{\rm even}}\simeq 10^4 \mu{\rm K}^4$ found in typical realizations. 
The  probabilities for the standard picture to match such  small values range from $10^{-4.3}$ to $10^{-2.8}$.
 }
\label{cumulative}
\end{figure*}
\subsubsection{Rank comparison  with standard realizations}

Our three comparisons are as follows. First, we generate $N = 2\cdot 10^6$ standard model realizations. We then evaluate $\Delta_{\text{\rm even}} (C(\Theta))$, $\Delta_{\text{minimal}} (C(\Theta))$, and $\Delta_{\text{maximal}} (C(\Theta))$ for these standard model realizations, such as those shown in Fig.\;(\ref{correlationspider}), as well as the different measured CMB maps. For a given residual $\Delta_{\text{\rm even}}$, $\Delta_{\text{minimal}}$, or $\Delta_{\text{maximal}}$, the variation of this residual for different measured CMB maps gives a measure of the sensitivity of the residual to Galactic model uncertainties.

The top panel of Fig.\;(\ref{cumulative}) shows cumulative probability, the fraction of standard realizations with $\Delta_{\text{\rm even}}$ smaller than the value shown on the horizontal axis, and the vertical lines show the values of $\Delta_{\text{\rm even}}$ for different measured CMB maps. 
These numbers confirm the visual impression from Fig.\;(\ref{correlationspider}): in the three {\sl Planck} maps the value of $\Delta_{\text{\rm even}}$ ranges from 1 to 9 $\mu{\rm K}^4$, compared with  values $\Delta_{\text{\rm even}}\simeq 10^4 \mu{\rm K}^4$  found in  typical  realizations.
Only a small fraction of standard  realizations come as close to zero $\Delta_{\text{\rm even}}$ as the  CMB temperature maps; the fraction in the {\sl Planck} maps   ranges from  $10^{-2.8}$ to   $10^{-4.3}$.

The middle panel shows the same quantities evaluated for $\Delta_{\text{minimal}}$, with the dipole-term adjustment described above. We again find that a small fraction of standard model realizations, ranging from about $10^{-1.8}$ (for {\sl WMAP}) to $10^{-2.8}$ (for Commander), come as close to zero $\Delta_{\text{minimal}}$ as the measured CMB temperature maps. The values of $C_1$ for the best-fit dipole for the maps  NILC, SMICA, COMMANDER, {\sl WMAP} without its monopole, and {\sl WMAP} with its monopole are approximately $365$, $341$, $322$, $392$, and $426$ $\mu$K$^2$, respectively.
 
The lower panel shows the same quantities evaluated for $\Delta_{\text{maximal}}$, with the dipole-term adjustment described above. We again find that a small fraction of standard model realizations, ranging from about $10^{-2.3}$ (for {\sl WMAP}) to $10^{-3.2}$ (for Commander), come as close to zero $\Delta_{\text{maximal}}$ as the measured CMB temperature maps. The values of $C_1$ for the best-fit dipole for the maps NILC, SMICA, COMMANDER, {\sl WMAP} without its monopole, and {\sl WMAP} with its monopole are approximately $384$, $390$, $389$, $440$, and $471$ $\mu$K$^2$, respectively.

The total residuals $\Delta_{\text{minimal}}$ and $\Delta_{\text{maximal}}$  are  significantly larger than the even residual $\Delta_{\text{\rm even}}$ over the narrower range of the 3D causal shadow.  However, the variation between the maps is also larger,  again consistent with the interpretation that departures from zero are due to inaccurate models of the Galaxy, and  that intrinsic CMB correlations actually vanish.

To evaluate the sensitivity of this comparison to the chosen cutoff $\ell_{\text{max}}$, we repeated them for every $\ell_{\text{max}}$ value ranging from $25$ to $35$. For each value, the significance of our results, i.e. the fraction of standard model realizations having a residual as low as that of measured CMB temperature maps, changed only slightly, less by at least an order of magnitude than the variations in significance between different maps.

\subsection{Interpretation}
\label{interpretation}

In standard inflation,  correlation varies widely among different realizations, and the tiny measured correlation must be interpreted as a statistical anomaly. Our rank   comparison (Fig.\;\ref{cumulative}) shows that the sky agrees with zero better than almost all standard realizations. The most direct comparison, as well as the smallest measured values of correlation, appear in  even-parity correlation.


Another interpretation is that the nearly-zero correlation is due to initial conditions more symmetric than generally assumed.
One possibility is an exact fundamental causal symmetry   that is not  included in the standard model. The measured angular range of minimal correlation agrees with a range of  zero correlation derived here  from 4D causal coherence. In this interpretation, the  measured departures from zero correlation are attributed to measurement error, dominated by contamination by the Galaxy. This  view is consistent with measured variation among the different maps.


\section{Conclusion}
\label{conclusion}

A surprisingly small absolute value of the large-angle CMB correlation function has been known since the first measurements with {\sl COBE}.
Subsequent measurements from {\sl WMAP}  and then {\sl Planck}  showed values successively closer to zero.  They are not generally thought to present a  compelling challenge to standard cosmological theory, both because the standard theory occasionally produces such small correlations by chance, and because there has not been a precisely formulated and physically compelling alternative expectation.

The significant new fact reported in this paper is that when the removed dipole component is accounted for, the magnitude of directly measured
correlation  over a range of angles around $\Theta=90^\circ$   
is  much smaller than previously documented.
Allowing for measurement uncertainty,  even-parity correlation  in a geometrically-calculated range of angles is consistent with zero, and at least several orders of magnitude smaller than expected from standard initial conditions. The probability of  correlations as small as those measured in  {\sl Planck} maps is $10^{-2.8}$ to  $10^{-4.3}$.

We are thus led to suspect that nearly-zero large angle correlation may not be an
accident of our particular sky, but  a  signature of  physical symmetry in cosmic initial conditions.
A range of zero correlation that matches the data  can be calculated geometrically
 from a causal  bound on the coherence of gravitational vacuum fluctuation states, which is not included in the standard theory of inflationary perturbations based on QFT.
 

As discussed in the Appendix, a symmetry of this kind is broadly consistent with  tests of classical concordance cosmology, which mainly (if not entirely) depend only on a nearly scale-invariant initial 3D power spectrum of perturbations averaged over large volumes. It is  also consistent with all other experimental tests of QFT, none of which depend on quantized  gravity.


If the causal-symmetry hypothesis is not true, it can be falsified by more precise measurement of nonzero correlations on the cosmic horizon within predicted causal shadows. 
The precision of the results reported here, and the significance of our null tests,  are not limited by any fundamental source of noise, but by the accuracy of  models  of Galactic foreground emission.
Tests of the symmetry could be improved with all-sky models of emission from the Galaxy  that allow more accurate measurement of the true CMB pattern on the largest scales. 
Other unique cosmological signatures of causal coherence are addressed briefly in the Appendix.



\section*{Data Availability}
 For data access from both {\sl WMAP} and {\sl Planck} \citep{https://doi.org/10.26131/irsa561,https://doi.org/10.26131/irsa558}, 
 we acknowledge use of the Legacy Archive for Microwave
Background Data Analysis (LAMBDA),
part of the High
Energy Astrophysics Science Archive Center (HEASARC), and the
NASA/ IPAC Infrared Science Archive, which is operated by
the Jet Propulsion Laboratory, California Institute of Technology, under contract with the National Aeronautics and Space
Administration.
\section*{Acknowledgements}
O.K. thanks the Cardiff Gravity Exploration Institute for its hospitality and acknowledges support from the UKRI Science and Technology Facilities Council (STFC) under grant ST/Y005082/1. We are also grateful for support from an anonymous donor.


\bibliographystyle{mnras}
\bibliography{nearly}

\begin{thebibliography}{}
\makeatletter
\relax
\def\mn@urlcharsother{\let\do\@makeother \do\$\do\&\do\#\do\^\do\_\do\%\do\~}
\def\mn@doi{\begingroup\mn@urlcharsother \@ifnextchar [ {\mn@doi@}
  {\mn@doi@[]}}
\def\mn@doi@[#1]#2{\def\@tempa{#1}\ifx\@tempa\@empty \href
  {http://dx.doi.org/#2} {doi:#2}\else \href {http://dx.doi.org/#2} {#1}\fi
  \endgroup}
\def\mn@eprint#1#2{\mn@eprint@#1:#2::\@nil}
\def\mn@eprint@arXiv#1{\href {http://arxiv.org/abs/#1} {{\tt arXiv:#1}}}
\def\mn@eprint@dblp#1{\href {http://dblp.uni-trier.de/rec/bibtex/#1.xml}
  {dblp:#1}}
\def\mn@eprint@#1:#2:#3:#4\@nil{\def\@tempa {#1}\def\@tempb {#2}\def\@tempc
  {#3}\ifx \@tempc \@empty \let \@tempc \@tempb \let \@tempb \@tempa \fi \ifx
  \@tempb \@empty \def\@tempb {arXiv}\fi \@ifundefined
  {mn@eprint@\@tempb}{\@tempb:\@tempc}{\expandafter \expandafter \csname
  mn@eprint@\@tempb\endcsname \expandafter{\@tempc}}}

\bibitem[\protect\citeauthoryear{Bardeen}{Bardeen}{1980}]{Bardeen1980}
Bardeen J.~M.,  1980, \mn@doi [Phys. Rev. D] {10.1103/PhysRevD.22.1882}, 22,
  1882

\bibitem[\protect\citeauthoryear{Baumann}{Baumann}{2011}]{Baumann2009}
Baumann D.,  2011, in {Physics of the large and the small, TASI 09}. pp
  523--686 (\mn@eprint {arXiv} {0907.5424}),
  \mn@doi{10.1142/9789814327183_0010}, \url
  {https://inspirehep.net/record/827549/files/arXiv:0907.5424.pdf}

\bibitem[\protect\citeauthoryear{{Bennett} et~al.,}{{Bennett}
  et~al.}{1994}]{Bennett1994}
{Bennett} C.~L.,  et~al., 1994, \mn@doi [\apj] {10.1086/174918}, \href
  {https://ui.adsabs.harvard.edu/abs/1994ApJ...436..423B} {436, 423}

\bibitem[\protect\citeauthoryear{Bennett et~al.,}{Bennett
  et~al.}{2011}]{WMAPanomalies}
Bennett C.~L.,  et~al., 2011, \mn@doi [The Astrophysical Journal Supplement
  Series] {10.1088/0067-0049/192/2/17}, 192, 17

\bibitem[\protect\citeauthoryear{Cohen, Kaplan  \& Nelson}{Cohen
  et~al.}{1999}]{CohenKaplanNelson1999}
Cohen A.~G.,  Kaplan D.~B.,   Nelson A.~E.,  1999, \mn@doi [Phys. Rev. Lett.]
  {10.1103/PhysRevLett.82.4971}, 82, 4971

\bibitem[\protect\citeauthoryear{Copi, Huterer, Schwarz  \& Starkman}{Copi
  et~al.}{2007}]{Copi2006}
Copi C.,  Huterer D.,  Schwarz D.,   Starkman G.,  2007, \mn@doi [Phys. Rev. D]
  {10.1103/PhysRevD.75.023507}, 75, 023507

\bibitem[\protect\citeauthoryear{Copi, Huterer, Schwarz  \& Starkman}{Copi
  et~al.}{2009}]{Copi2008}
Copi C.~J.,  Huterer D.,  Schwarz D.~J.,   Starkman G.~D.,  2009, \mn@doi [Mon.
  Not. Roy. Astron. Soc.] {10.1111/j.1365-2966.2009.15270.x}, 399, 295

\bibitem[\protect\citeauthoryear{{Copi}, {Huterer}, {Schwarz}  \&
  {Starkman}}{{Copi} et~al.}{2015}]{Copi2015}
{Copi} C.~J.,  {Huterer} D.,  {Schwarz} D.~J.,   {Starkman} G.~D.,  2015,
  \mn@doi [\mnras] {10.1093/mnras/stv1143}, \href
  {https://ui.adsabs.harvard.edu/abs/2015MNRAS.451.2978C} {451, 2978}

\bibitem[\protect\citeauthoryear{Copi, O'Dwyer  \& Starkman}{Copi
  et~al.}{2016}]{Copi2016}
Copi C.~J.,  O'Dwyer M.,   Starkman G.~D.,  2016, \mn@doi [Monthly Notices of
  the Royal Astronomical Society] {10.1093/mnras/stw2163}, 463, 3305

\bibitem[\protect\citeauthoryear{Copi, Gurian, Kosowsky, Starkman  \&
  Zhang}{Copi et~al.}{2019}]{Copi2018}
Copi C.~J.,  Gurian J.,  Kosowsky A.,  Starkman G.~D.,   Zhang H.,  2019,
  \mn@doi [Mon. Not. Roy. Astron. Soc.] {10.1093/mnras/stz2962}, 490, 5174

\bibitem[\protect\citeauthoryear{Ellis}{Ellis}{1999}]{Ellis1999}
Ellis G. F.~R.,  1999, \mn@doi [Classical and Quantum Gravity]
  {10.1088/0264-9381/16/12A/303}, 16, A37

\bibitem[\protect\citeauthoryear{Francis \& Peacock}{Francis \&
  Peacock}{2010}]{Francis2010}
Francis C.~L.,  Peacock J.~A.,  2010, \mn@doi [Monthly Notices of the Royal
  Astronomical Society] {10.1111/j.1365-2966.2010.16866.x}, 406, 14

\bibitem[\protect\citeauthoryear{Giar\`e, Di~Valentino  \& Melchiorri}{Giar\`e
  et~al.}{2024}]{Giare2023}
Giar\`e W.,  Di~Valentino E.,   Melchiorri A.,  2024, \mn@doi [Phys. Rev. D]
  {10.1103/PhysRevD.109.103519}, 109, 103519

\bibitem[\protect\citeauthoryear{Gorski, Hivon, Banday, Wandelt, Hansen,
  Reinecke  \& Bartelmann}{Gorski et~al.}{2005}]{Gorski_2005}
Gorski K.~M.,  Hivon E.,  Banday A.~J.,  Wandelt B.~D.,  Hansen F.~K.,
  Reinecke M.,   Bartelmann M.,  2005, \mn@doi [The Astrophysical Journal]
  {10.1086/427976}, 622, 759

\bibitem[\protect\citeauthoryear{Hagimoto, Hogan, Lewin  \& Meyer}{Hagimoto
  et~al.}{2020}]{Hagimoto_2020}
Hagimoto R.,  Hogan C.,  Lewin C.,   Meyer S.~S.,  2020, \mn@doi [The
  Astrophysical Journal] {10.3847/2041-8213/ab62a0}, 888, L29

\bibitem[\protect\citeauthoryear{Hinshaw, Banday, Bennett, G{\'{o}}rski, Kogut,
  Lineweaver, Smoot  \& Wright}{Hinshaw et~al.}{1996}]{Hinshaw_1996}
Hinshaw G.,  Banday A.~J.,  Bennett C.~L.,  G{\'{o}}rski K.~M.,  Kogut A.,
  Lineweaver C.~H.,  Smoot G.~F.,   Wright E.~L.,  1996, \mn@doi [The
  Astrophysical Journal] {10.1086/310076}, 464, L25

\bibitem[\protect\citeauthoryear{Hogan}{Hogan}{2019}]{Hogan2019}
Hogan C.,  2019, \mn@doi [Phys. Rev. D] {10.1103/PhysRevD.99.063531}, 99,
  063531

\bibitem[\protect\citeauthoryear{Hogan \& Meyer}{Hogan \&
  Meyer}{2022}]{Hogan_2022}
Hogan C.,  Meyer S.~S.,  2022, \mn@doi [Classical and Quantum Gravity]
  {10.1088/1361-6382/ac4829}, 39, 055004

\bibitem[\protect\citeauthoryear{Hollands \& Wald}{Hollands \&
  Wald}{2004}]{HollandsWald2004}
Hollands S.,  Wald R.~M.,  2004, \mn@doi [General Relativity and Gravitation]
  {10.1023/b:gerg.0000048980.00020.9a}, 36, 2595

\bibitem[\protect\citeauthoryear{Hou, Slepian  \& Cahn}{Hou
  et~al.}{2023}]{Hou2022}
Hou J.,  Slepian Z.,   Cahn R.~N.,  2023, \mn@doi [Mon. Not. Roy. Astron. Soc.]
  {10.1093/mnras/stad1062}, 522, 5701

\bibitem[\protect\citeauthoryear{Hu \& Dodelson}{Hu \&
  Dodelson}{2002}]{Hu:2001bc}
Hu W.,  Dodelson S.,  2002, \mn@doi [Ann. Rev. Astron. Astrophys.]
  {10.1146/annurev.astro.40.060401.093926}, 40, 171

\bibitem[\protect\citeauthoryear{{Ijjas} \& {Steinhardt}}{{Ijjas} \&
  {Steinhardt}}{2016}]{Ijjas2015}
{Ijjas} A.,  {Steinhardt} P.~J.,  2016, \mn@doi [Classical and Quantum Gravity]
  {10.1088/0264-9381/33/4/044001}, \href
  {https://ui.adsabs.harvard.edu/abs/2016CQGra..33d4001I} {33, 044001}

\bibitem[\protect\citeauthoryear{Kwon}{Kwon}{2025}]{Kwon2022}
Kwon O.,  2025, \mn@doi [Foundations of Physics] {10.1007/s10701-025-00827-4},
  55, 19

\bibitem[\protect\citeauthoryear{{Lewis} \& {Challinor}}{{Lewis} \&
  {Challinor}}{2011}]{CAMB2011}
{Lewis} A.,  {Challinor} A.,  2011, {CAMB: Code for Anisotropies in the
  Microwave Background}, Astrophysics Source Code Library, record ascl:1102.026
  (\mn@eprint {} {https://ascl.net/1102.026})

\bibitem[\protect\citeauthoryear{Muir, Adhikari  \& Huterer}{Muir
  et~al.}{2018}]{Muir2018}
Muir J.,  Adhikari S.,   Huterer D.,  2018, \mn@doi [Phys. Rev. D]
  {10.1103/PhysRevD.98.023521}, 98, 023521

\bibitem[\protect\citeauthoryear{Peebles}{Peebles}{2022}]{Peebles2022}
Peebles P. J.~E.,  2022, \mn@doi [Annals Phys.] {10.1016/j.aop.2022.169159},
  447, 169159

\bibitem[\protect\citeauthoryear{Penrose}{Penrose}{1989}]{Penrose1989}
Penrose R.,  1989, Annals of the New York Academy of Sciences, 571, 249

\bibitem[\protect\citeauthoryear{Philcox}{Philcox}{2022}]{Philcox2022}
Philcox O. H.~E.,  2022, \mn@doi [Phys. Rev. D] {10.1103/PhysRevD.106.063501},
  106, 063501

\bibitem[\protect\citeauthoryear{Philcox}{Philcox}{2023}]{Philcox2023}
Philcox O. H.~E.,  2023, \mn@doi [Phys. Rev. Lett.]
  {10.1103/PhysRevLett.131.181001}, p. 181001

\bibitem[\protect\citeauthoryear{{Planck Collaboration}}{{Planck
  Collaboration}}{2016}]{Planck2016}
{Planck Collaboration} 2016, \mn@doi [Astron. Astrophys.]
  {10.1051/0004-6361/201526681}, \href
  {https://ui.adsabs.harvard.edu/abs/2016A&A...594A..16P} {594, A16}

\bibitem[\protect\citeauthoryear{{Planck Collaboration}}{{Planck
  Collaboration}}{2020a}]{2020A&A...641A...6P}
{Planck Collaboration} 2020a, \mn@doi [Astron. Astrophys.]
  {10.1051/0004-6361/201833910}, \href
  {https://ui.adsabs.harvard.edu/abs/2020A&A...641A...6P} {641, A6}

\bibitem[\protect\citeauthoryear{{Planck Collaboration}}{{Planck
  Collaboration}}{2020b}]{Planck-2018-Isotropy}
{Planck Collaboration} 2020b, \mn@doi [Astron. Astrophys.]
  {https://doi.org/10.1051/0004-6361/201935201}, 641, A7

\bibitem[\protect\citeauthoryear{{Planck Collaboration}}{{Planck
  Collaboration}}{2020c}]{Akrami:2019bkn}
{Planck Collaboration} 2020c, \mn@doi [Astron. Astrophys.]
  {10.1051/0004-6361/201935201}, 641, A7

\bibitem[\protect\citeauthoryear{{Planck Team}}{{Planck
  Team}}{2013}]{https://doi.org/10.26131/irsa561}
{Planck Team} 2013, Planck Public Data Release 1 External Data,
  \mn@doi{10.26131/IRSA561}, \url
  {https://catcopy.ipac.caltech.edu/dois/doi.php?id=10.26131/IRSA561}

\bibitem[\protect\citeauthoryear{{Planck Team}}{{Planck
  Team}}{2020}]{https://doi.org/10.26131/irsa558}
{Planck Team} 2020, Planck Public Data Release 3 Maps,
  \mn@doi{10.26131/IRSA558}, \url
  {https://catcopy.ipac.caltech.edu/dois/doi.php?id=10.26131/IRSA558}

\bibitem[\protect\citeauthoryear{{Sachs} \& {Wolfe}}{{Sachs} \&
  {Wolfe}}{1967}]{Sachs1967}
{Sachs} R.~K.,  {Wolfe} A.~M.,  1967, \mn@doi [\apj] {10.1086/148982}, \href
  {https://ui.adsabs.harvard.edu/abs/1967ApJ...147...73S} {147, 73}

\bibitem[\protect\citeauthoryear{{Spergel} et~al.,}{{Spergel}
  et~al.}{2003}]{Spergel2003}
{Spergel} D.~N.,  et~al., 2003, \mn@doi [\apjs] {10.1086/377226}, \href
  {https://ui.adsabs.harvard.edu/abs/2003ApJS..148..175S} {148, 175}

\bibitem[\protect\citeauthoryear{Stamp}{Stamp}{2015}]{Stamp_2015}
Stamp P. C.~E.,  2015, \mn@doi [New Journal of Physics]
  {10.1088/1367-2630/17/6/065017}, 17, 065017

\bibitem[\protect\citeauthoryear{Vermeulen, Aiello, Ejlli, Griffiths, James,
  Dooley  \& Grote}{Vermeulen et~al.}{2021}]{vermeulen2020}
Vermeulen S.,  Aiello L.,  Ejlli A.,  Griffiths W.,  James A.,  Dooley K.,
  Grote H.,  2021, \mn@doi [Classical and Quantum Gravity]
  {10.1088/1361-6382/abe757}, 38, 085008

\bibitem[\protect\citeauthoryear{Vermeulen et~al.,}{Vermeulen
  et~al.}{2025}]{vermeulen2024photon}
Vermeulen S.~M.,  et~al., 2025, \mn@doi [Phys. Rev. X]
  {10.1103/PhysRevX.15.011034}, 15, 011034

\bibitem[\protect\citeauthoryear{Vilasini \& Renner}{Vilasini \&
  Renner}{2024}]{Vilasini2024}
Vilasini V.,  Renner R.,  2024, \mn@doi [Phys. Rev. Lett.]
  {10.1103/PhysRevLett.133.080201}, 133, 080201

\bibitem[\protect\citeauthoryear{Weinberg}{Weinberg}{2008}]{Weinberg2008}
Weinberg S.,  2008, {Cosmology}.
Oxford University Press, \url
  {http://www.oup.com/uk/catalogue/?ci=9780198526827}

\bibitem[\protect\citeauthoryear{Zeilinger}{Zeilinger}{1999}]{Zeilinger1999}
Zeilinger A.,  1999, \mn@doi [Rev. Mod. Phys.] {10.1103/RevModPhys.71.S288},
  71, S288

\makeatother
\end{thebibliography}
\section{Appendix}
\label{appendix}

\begin{figure*}
\begin{centering} \includegraphics[width=.9\linewidth]{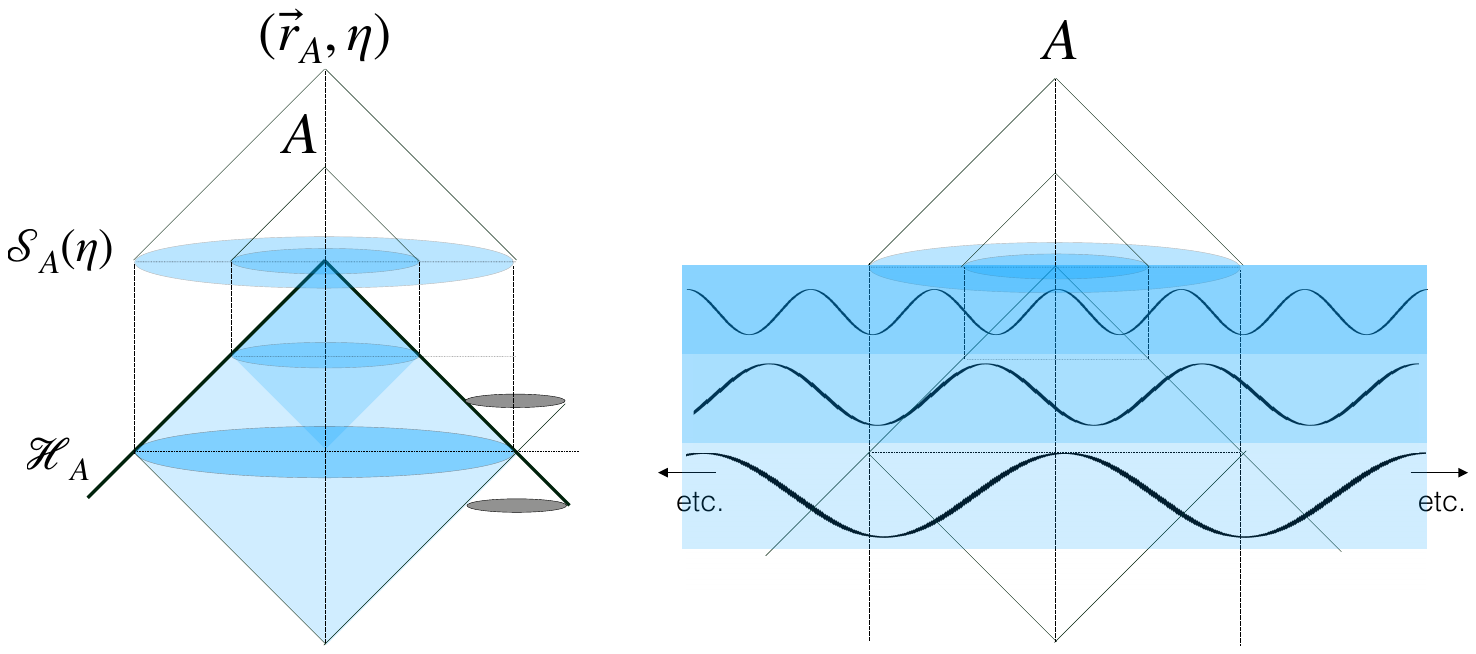}
\par\end{centering}
\protect\caption{Geometrical structures of coherent quantum states in the causally coherent picture and the standard QFT picture. The left panel shows a conformal causal diagram of the history of a comoving region around world line $A$, as in Fig. (\ref{nested}). The comoving footprint  ${\cal S}_A(\eta)$ of the horizon  ${\cal H}_A$ bounds the 4D  coherent fluctuations of a causal diamond that generates  perturbations correlated with  world line $A$ at $(\vec r_A, \eta)$.   At right, the same  causal structure is shown with a sketch of the spatial distribution  of three spacelike-coherent wave modes in the standard QFT picture, as their oscillations freeze during inflation. In this model, the  final frozen spatial 3D pattern  is fixed by a  particular configuration of coherent modes that extend far beyond the horizon, as sketched here.  The particular pattern is determined by the initial conditions of all modes specified
in the initial vacuum state, over a  spacelike region much larger than the size of the inflationary horizon when they freeze. This model builds in spacelike correlations on scales much larger than ${\cal H}_A$ for any value of $\eta$, and  generates considerable variance in realizations of large-angle anisotropy.  }
\label{diamonds}
\end{figure*}
\subsection{Causal structure  and initial conditions in the QFT model}

The standard QFT model of inflationary perturbations \citep{Weinberg2008}  starts with the established quantum physics of fields and  the established classical theory of space-time, and combines and extrapolates them  into a new physical regime.
It is possible that it introduces incorrect assumptions about the initial state of the system, and perhaps also about behavior of quantum gravity, in particular about the geometrical structure of  coherent quantum states of geometry in four dimensions.

It is useful to review  basic  assumptions of standard inflation theory for the initial state and the evolution of the system, and contrast these with a causally-coherent picture.
These contrasting models of initial conditions are  based on different models of gravitational fluctuations and their conversion into classical perturbations. They ultimately depend on how    nonlocal quantum phenomena influence   causal relationships between events, and  how locality and causality emerge from a quantum system, which remain  unsolved problems in quantum gravity \citep{Stamp_2015}.
The two alternatives are illustrated in Fig. (\ref{diamonds}). 

In the standard QFT inflation model, 
the quantum state of the system is described relative  to a perfectly uniform  classical background universe that encompasses some large initial patch. The patch need  not be infinitely large, but must be much larger than the current CMB horizon scale. 
Scalar inhomogeneities around this uniform background are decomposed into comoving Fourier modes of quantum fields.

In a general field vacuum state,  each mode is  in a ground state with zero mean amplitude but a nonzero mean square amplitude due to zero-point fluctuations.
The  field pattern in this state is determined by the phases of zero point oscillations of all the modes.
In general, the field value at any particular position is an indeterminate superposition of possible values, like the position of a quantum particle in a state prepared as a wave.

For standard inflation theory, the vacuum state for a particular universe is initially specified to be in a state with definite phases  for every mode. These   random numbers  ultimately specify the  realized pattern of perturbations.  The entire initial patch is in this definite state, so the quantum system can be said to be already ``collapsed'' in the initial conditions--- that is, not in a superposition.  A complete wave function of the initial vacuum would be a superposition of all the standard realizations. Of course, we live in just one of these, in which  all of the mode phases are assumed to have been coherently evolving since they were laid down initially.  

The accelerating expansion converts quantum fluctuations into classical gravitational perturbations.
In the standard picture, vacuum fluctuations are said to ``freeze'' into their final configuration  when their wavelength approximately matches the scale of the inflationary horizon ${\cal H}$.
 The
freezing of each mode is controlled by a wave equation.  Coherent oscillations of each comoving  field mode cool coherently by cosmic expansion into a  classical configuration of constant curvature perturbation at each location, determined entirely by its initial phase and amplitude. 
The global perturbation pattern represented by  each frozen mode is interpreted as a classical curvature perturbation, which generates perturbations with correlations at  spacelike separation over the entire mode, extending over the whole initial patch.  As far as predictions of the model are concerned, the coherent quantum state and its correlations are spatially unbounded.

Thus, the conversion of quantum vacuum  fluctuations into classical curvature perturbations is modeled as a gradual expansion-driven cooling of randomly initialized coherent standing plane waves.
 In this sense, freezing does not actually describe a quantum-to-classical conversion:  the model assumes that the quantum state is collapsed into a classical state already over the entire initial patch when initial conditions are laid down. There is no part of  history after the initial state during which the metric is in a superposition of different possibilities.  That is why the spatial pattern of a classical realization, such as those used in the rank comparisons above, is determined entirely by the set of random mode phases  specified in the initial vacuum state.

 In QFT, the independent, spatially-coherent  elements of the quantum system are the modes, which have a comoving size far exceeding their wavelength, and remain coherent throughout inflation.
 If the causal coherence hypothesis is correct, this model does
  not correctly account for causal quantum relationships on scales comparable to or larger than  horizons.
The QFT  approximation omits effects introduced into a quantum state by   inflationary horizons,  the incoming spherical null surfaces that terminate on  each world line at the end of inflation.

To take one example, an inflationary horizon  defines a one-way boundary of  causal relationships with its world line: information only passes through it in the outwards radial direction, in the same way that information only passes radially inwards at a black hole horizon.
This asymmetry is not modeled by  the unbounded coherence  assumed in the standard picture, which assumes a coherent superposition of opposite propagation directions in  standing plane waves, in order to  result in zero total momentum in the frame of the initial background patch.

More generally,  coherent plane waves on spacelike surfaces do not conform with  relational geometrical causal structures in space-time. An actual physical  horizon is a sharp physical causal boundary on a spherical null surface converging on a particular world line, which is not  planar, wavelike or spacelike. Physical vacuum fluctuation states that conform to this structure  entangle  in  ways that are not accounted for in the standard inflation picture. 

Suppose instead  that causally coherent  quantum fluctuation states are confined within  causal diamonds. The correlations they generate have compact footprints that do not extend beyond  horizons. 
Fluctuations freeze into relational classical perturbations only when world lines cross inflationary horizons.
Unlike the initial conditions in the QFT model, perturbations remain in an indeterminate quantum superposition within horizons.
This hypothesis allows
quantum fluctuations to create correlations of   classical perturbations between world lines only as far as their entanglement within physical causal boundaries, that is, actual horizons.  

An observable relational potential difference between locations has contributions from perturbation and from the curvature of the initial classical background. Although we have used the usual convention of perturbations defined on a uniform classical background, in a causally-coherent picture the background space-time is not separately predefined as it is in the standard picture; it emerges relationally from a quantum system along with  perturbations, and is only fixed at the end of inflation. 



In  standard QFT inflation, orthogonal components of momenta are assumed  to commute thoughout inflation; hence, projections of field modes along each axis are  separable  quantum systems. Independent Gaussian perturbations in 3D generate  independent Gaussian angular harmonics,  which leads to the standard cosmic variance for realized classical angular correlation.
As seen from the realizations shown  in Fig. (\ref{correlationspider}), a causal shadow is incompatible with standard cosmic variance, which predicts vanishing angular correlation only for a set of angular separations of measure zero.
As explained above, symmetries of angular correlations  can   appear if  perturbations   entangle nonlocally with causal structure on the scale of the horizon, so different directions are no longer separable \citep{Hogan2019,Hogan_2022}. 

\subsection{Modifications of concordance  cosmology}

Causal coherence  significantly modifies some cosmological inferences and  projections derived from the field dynamics of the QFT model, such as the spectrum of primordial gravitational waves (cosmological tensor modes), or the relationship of the effective inflaton potential to the scalar fluctuation spectrum. 
However, most current tests  of $\Lambda$CDM cosmology depend mainly on the 3D power spectrum or two-point correlation function of  curvature perturbations averaged over all directions in a large volume.   Standard inflation theory produces the required nearly-scale-invariant 3D power spectrum,  given a suitably tuned effective inflaton potential, but that spectrum is not unique to QFT; the same  spectrum would also be produced by causally-coherent fluctuations whose  variance is determined by a suitably slowly-changing physical horizon radius. The main features of standard post-inflation cosmology are the same in the two cases. 

Some standard cosmological predictions that depend on statistical isotropy and independence of modes in $\vec k$ space would be modified by   higher order 3D correlations of causally coherent perturbations. 
These occur on  comoving scales smaller than the current CMB surface. 


 For example, a significant systematic modification is expected for large-angle polarization anisotropy from the epoch of reionization, which impacts some estimates of the optical depth\footnote{We are grateful to G. Holder for bringing this situation to our attention.}.  Suppose that the total quadrupole moment  $C_{2}$ of the CMB viewed by electrons at reionization, on their different and smaller horizons, is the same as that of the CMB today,  which is about a factor of  four less than the standard expectation.  This  reduces the low-$\ell$ reionization bump in the polarization ($EE$) spectrum for a given optical depth $\tau$, so canonical {\sl Planck} estimates of optical depth, determined mainly by the low-$\ell$ $EE$ bump amplitude, are significantly lower than the true value.
  The required true optical depth increases by  approximately the square root of the ratio of the standard expected quadrupole coefficient $C_2$  to the true value, so instead of the usual $EE$-estimated {\sl Planck} value $\tau\simeq .05$, the optical depth required to agree with the same {\sl Planck} $EE$ measurement increases by up to a factor of two.  A higher optical depth  improves the overall consistency of the flat $\Lambda$CDM  model with   measurements, including {\sl Planck} spectra  at  $\ell>30$ \citep{Giare2023}.

Exotic higher-order correlations  might also  be directly measured  in  large-volume spectroscopic surveys, such as {\sl BOSS}, {\sl DESI}, and {\sl Euclid} \citep{Hogan2019}.
A particularly distinctive signature  could appear as global parity violation in the 3D mass distribution.  At very large angular  separations,  the CMB  correlation function is nonzero and negative,
    $C(\Theta\rightarrow \pi)<0$,
which also manifests as an excess of odd- over even- parity spectral perturbation power  measured in  CMB power spectra to $\ell\simeq 30$  \citep{Planck2016}. 
This  parity violation, if it is attributed to  universal causal coherence,  should affect  perturbations on all linear scales in 3D. 
Such exotic  parity violation may account for recent detections of parity violation in the large-scale galaxy distribution
 \citep{Hou2022,Philcox2022} that are difficult to account for in the standard scenario with QFT-based $\mathbb{P}$-symmetry violation \citep{Philcox2023}.

\end{document}